\documentclass[aps,twocolumn,pra,showpacs]{revtex4-1}
\usepackage{graphicx,color,textcomp}
\usepackage{mathrsfs,amsmath}   
\usepackage{psfrag,overpic,pst-all}
\usepackage{amssymb}

\newcommand{\rme}{{\mathrm e}}

\begin{document}

\title{Frequency comb formation in doubly resonant second-harmonic generation}

\author{F. Leo$^1$}
\email{f.leo@auckland.ac.nz}
\author{T. Hansson$^{2,3}$}
\author{I. Ricciardi$^4$}
\author{M. De~Rosa$^4$}
\author{S. Coen$^1$}
\author{S. Wabnitz$^{3,4}$}
\author{M. Erkintalo$^1$}
\email{m.erkintalo@auckland.ac.nz}

\affiliation{$^1$The Dodd-Walls Centre for Photonic and Quantum Technologies, Department of Physics, The University of Auckland, Auckland 1142, New Zealand}
\affiliation{$^2$Department of Applied Physics, Chalmers University of Technology, SE-41296 G\"oteborg, Sweden}
\affiliation{$^3$Dipartimento di Ingegneria dell'Informazione, Universit\`a di Brescia, via Branze 38, 25123 Brescia, Italy}
\affiliation{$^4$CNR-INO, Istituto Nazionale di Ottica, Via Campi Flegrei 34, 80078 Pozzuoli (NA), Italy}

\begin{abstract}
{We theoretically study the generation of optical frequency combs and corresponding pulse trains in doubly resonant intracavity second-harmonic generation (SHG). We find that, despite the large temporal walk-off characteristic of realistic cavity systems, the nonlinear dynamics can be accurately and efficiently modelled using a pair of coupled mean-field equations. Through rigorous stability analysis of the system's steady-state continuous wave solutions, we demonstrate that walk-off can give rise to a new, previously unexplored regime of temporal modulation instability (MI). Numerical simulations performed in this regime reveal rich dynamical behaviours, including the emergence of temporal patterns that correspond to coherent optical frequency combs. We also demonstrate that the two coupled equations that govern the doubly resonant cavity behaviour can, under typical conditions, be reduced to a single mean-field equation akin to that describing the dynamics of singly resonant cavity SHG [F. Leo et al., Phys. Rev. Lett. 116, 033901 (2016)]. This reduced approach allows us to derive a simple expression for the MI gain, thus permitting to acquire significant insight into the underlying physics. We anticipate that our work will have wide impact on the study of frequency combs in emerging doubly resonant cavity SHG platforms, including quadratically nonlinear microresonators.}
\end{abstract}

\maketitle

\section{Introduction}
Continuously-driven Kerr microresonators have recently emerged as an attractive platform for the generation of optical frequency combs and corresponding high repetition rate pulse trains~\cite{kippenberg_microresonator_2011}. In these devices, comb formation proceeds via the third-order optical ``Kerr'' nonlinearity~\cite{delhaye_optical_2007, chembo_spectrum_2010}. Modulation instability (MI) driven by parametric four-wave mixing leads to the generation of signal and idler sidebands that subsequently interact so as to form an array of equidistant spectral components~\cite{hansson_dynamics_2013}. For suitable parameters, the phases of the comb lines can lock ~\cite{ferdous_spectral_2011, herr_universal_2012, saha_modelocking_2013}, thus leading to the formation of stable pulse sequences that may correspond to periodic MI patterns~\cite{coen_universal_2013} or localized temporal cavity solitons~\cite{coen_modeling_2013, herr_temporal_2014, brasch_photonic_2016, yi_generation_2015}. Before their observation in microresonators, similar dynamics and temporal structures were discovered and investigated in macroscopic optical fibre cavities~\cite{haelterman_dissipative_1992, coen_continuous-wave_2001, leo_temporal_2010, leo_dynamics_2013, jang_ultraweak_2013}; analogous phenomena in spatial nonlinear optics are rooted into an even longer history~\cite{mclaughlin_solitary_1983,mclaughlin_new_1985,lugiato_spatial_1987, ackemann_fundamentals_2009, barland_cavity_2002}.

Whilst Kerr nonlinearity underpins the dynamics observed in all \emph{microresonator} comb studies reported to date, recent experiments in \emph{bulk free-space} cavities have remarkably demonstrated that frequency combs can also arise purely through second-order $\chi^{(2)}$ nonlinear effects~\cite{ulvila_frequency_2013, ulvila_high-power_2014, ricciardi_frequency_2015}. Motivated by these experimental findings, we  recently introduced a general theoretical framework that allows for the temporal and spectral dynamics of light in such quadratically nonlinear resonators to be described, both analytically and numerically~\cite{leo_walkoff_2015}. To facilitate comparison with experiments, in~\cite{leo_walkoff_2015} we focussed on singly resonant, phase-matched cavity-enhanced second-harmonic generation (SHG), obtaining good agreement with the corresponding measurements~\cite{ricciardi_frequency_2015}. Our theoretical analysis showed that the temporal walk-off between the fields at the fundamental and second-harmonic frequencies plays a key role in the frequency comb formation dynamics, underpinning the MI process and the subsequent formation of drifting pulse trains.

\begin{figure*}[htb]
 \centering
 \includegraphics[width = 14.5cm, clip=true]{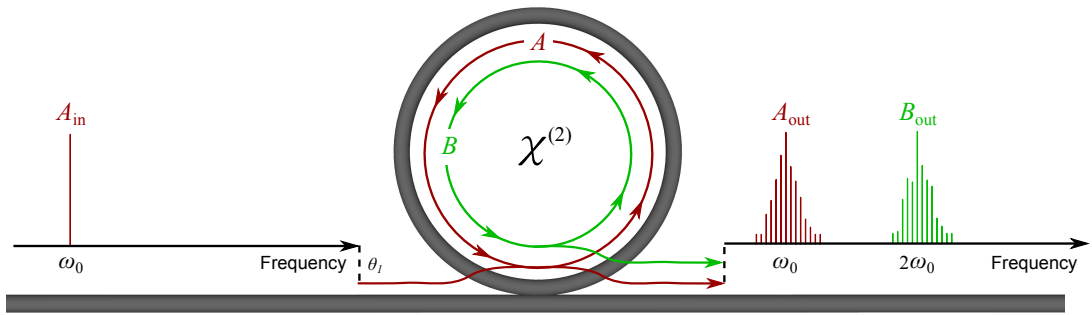}
\caption{Schematic example of the system under study. A ring resonator with a $\chi^{(2)}$ nonlinearity is driven with a cw field $A_\mathrm{in}$ at $\omega_0$. Phase-matched SHG gives rise to a field $B$ with carrier frequency $2\omega_0$ that resonates together with the fundamental field $A$. The quadratically nonlinear interaction of the fields may result in the generation of frequency combs around $\omega_0$ and $2\omega_0$.} \label{schematic}
\end{figure*}

Although purely quadratic frequency combs have so far only been observed in bulk free-space cavities~\cite{ulvila_frequency_2013, ulvila_high-power_2014, ricciardi_frequency_2015}, several platforms have recently emerged that may allow for similar sources to be implemented with integrated microresonators~\cite{furst_naturally_2010, lin_wide_2013, lin_continuous_2014, kuo_second_2014, mariani_2014,  vukovic_tunable_2015, mu_algas_2015}. In such devices, all interacting fields are likely to simultaneously resonate and exhibit low-losses, in stark contrast with the singly resonant system studied in~\cite{ricciardi_frequency_2015, leo_walkoff_2015}, where the second-harmonic field was fully extracted from the cavity after each round trip. Previous studies in \emph{spatially diffractive} $\chi^{(2)}$ resonators have established that the build-up and evolution of the cavity fields may differ considerably depending on the number of resonating fields~\cite{lodahl_thesis}, prompting one to anticipate that the particular \emph{singly resonant} mean-field equation derived in~\cite{leo_walkoff_2015} will not be directly suitable for the modelling of \emph{doubly resonant} microresonator SHG systems. Although the dynamics of doubly resonant cavity SHG has been extensively investigated in the past~\cite{mcneil_self_1978, drummond_non-equilibrium_1980, schiller_quadruply_1993, marte_competing_1994, lodahl_spatiotemporal_2001}, the vast majority of studies have only considered a very limited number of spectral components (up to 4). Such an approach is unsuitable for the modelling of frequency combs with hundreds or even thousands of spectral components. In 1996, Trillo and Haelterman introduced a full time-domain model of cavity SHG~\cite{trillo_pulse-train_1996}, predicting the formation of pulse trains via MI. Their approach does not suffer from limitations on the number of frequency components; however, the analysis presented in~\cite{trillo_pulse-train_1996} was restricted to negligible temporal walk-off, which is not a suitable condition for realistic quadratic materials~\cite{leo_walkoff_2015}. Doubly resonant cavity-enhanced SHG has also been studied in the context of spatial pattern formation~\cite{etrich_pattern_1997, etrich_solitary_1997}, but those studies also offer limited insights into dispersive resonators because spatial systems do not have a counterpart for \emph{strong} temporal walk-off. Given the growing interest in dispersive $\chi^{(2)}$ microresonators~\cite{furst_naturally_2010, lin_wide_2013, lin_continuous_2014, kuo_second_2014, mariani_2014, vukovic_tunable_2015, mu_algas_2015}, there is clearly a need for realistic models capable of predicting the associated nonlinear optical phenomena.

In this Article, we theoretically examine the full temporal and spectral dynamics of intracavity SHG when the fundamental and the second harmonic fields are both resonant, focussing in particular our attention on the previously unexplored role of temporal walk-off. Our main intent is to present model equations that allow for the simulation and analysis of such cavity systems, as well as to describe some general features and illustrative predictions derived from that model; a full analysis of the bifurcations and associated temporal dynamics is nevertheless left for future work. Our study shows that, even for a large temporal walk-off, doubly resonant cavity SHG can be efficiently and accurately modelled by using two coupled mean-field equations. Although the system's general nonlinear behaviour differs considerably from the singly resonant configuration studied in~\cite{leo_walkoff_2015}, we find that, similarly to that configuration, the temporal walk-off between the fundamental and the second-harmonic fields plays a very significant role. Indeed, we find that the large walk-off values of realistic quadratic cavities can give rise to new MI bands that display completely different features with respect to those identified previously~\cite{etrich_pattern_1997, trillo_pulse-train_1996} for negligible walk-off. Numerical simulations performed in this regime reveal very rich dynamical behaviours, including the formation of stable temporal patterns corresponding to coherent optical frequency combs. We also show that the two coupled equations that govern the dynamics of the doubly resonant system can be reduced, under realistic approximations, to a single mean-field equation that displays formal similarities with the corresponding equation that governs singly resonant cavity SHG. The significantly different dynamics that manifest themselves in the two types of cavities are simply captured by different nonlinear response functions. Analysis of the single reduced mean-field equation allows us to derive a simple expression for the system's MI gain spectrum, thus enabling us to obtain considerable physical insights into the underlying mechanisms. We expect that our theoretical framework and results will be highly relevant in the study of frequency comb generation in emerging quadratically nonlinear cavities.

\section{Coupled mean-field equations}
\label{models}
A schematic of the physical system under study is shown in Fig.~\ref{schematic}. We consider slowly varying electric field envelopes $A$ and $B$ with carrier frequencies $\omega_0$ and $2\omega_0$, respectively, circulating in a dispersive, quadratically nonlinear ring resonator that is driven with a continuous wave (cw) field $A_\mathrm{in}$ at the fundamental frequency $\omega_0$. We assume that diffraction can be neglected, and that the SHG process $\omega_0+\omega_0 = 2\omega_0$ is the dominant nonlinear interaction (i.e., down-conversion to the vicinity of $\omega_0/2$ as well as nonlinearities beyond quadratic are assumed highly phase-mismatched or otherwise negligible). Furthermore, we assume that the resonator exhibits high finesse both at the fundamental and the second-harmonic frequencies, so that the slowly varying envelopes do not vary significantly over a single round trip. Analogously to the case of Kerr nonlinear resonators~\cite{haelterman_dissipative_1992,coen_modeling_2013}, the infinite-dimensional map introduced in~\cite{leo_walkoff_2015} (see also~\cite{buryak_optical_2002}) can under these conditions be reduced to two coupled mean-field equations (see Appendix~\ref{Apx1} for full details). To facilitate comparison with earlier studies in spatially diffractive systems~\cite{etrich_pattern_1997, etrich_solitary_1997}, we write these equations in dimensionless form (dimensional forms are given in Appendix~\ref{Apx1}):
\begin{align}
&\frac{\partial v_1}{\partial t} = \left[-1 - {i}\Delta_1 -{i}\eta_1\frac{\partial^2}{\partial \tau^2}\right]v_1+ {i} \rme^{-{i}\xi}\mathrm{sinc}(\xi)v_2v_1^* +S\label{MF1}\\
&\frac{\partial v_2}{\partial t} = \left[-\alpha - {i}\Delta_2-d\frac{\partial}{\partial \tau} -{i}\eta_2\frac{\partial^2}{\partial \tau^2}\right]v_2 + {i} \rme^{{i}\xi}\mathrm{sinc}(\xi) v_1^2.\label{MF2}
\end{align}
Here $v_{1,2}(t,\tau)$ are the normalised slowly varying field envelopes at the fundamental and second-harmonic frequencies, respectively, $t$ is a ``slow time'' variable that describes field evolution on the time scale of the cavity photon lifetime, and $\tau$ is a ``fast time'' variable that allows for the temporal profiles of the fields to be described. Note that the periodic nature of the resonator imposes the condition $v_{1,2}(t,\tau)=v_{1,2}(t,\tau+\tau_\mathrm{s})$, where $\tau_\mathrm{s}$ is the normalized cavity round-trip time (at the fundamental frequency). The first two terms on the right hand side of Eqs.~\eqref{MF1} and ~\eqref{MF2} account for the cavity losses and the cavity phase detuning, respectively, while the second-order fast time ($\tau$) derivatives denote group-velocity dispersion (GVD). In addition, the terms proportional to $\mathrm{sinc}(\xi)$ describe the quadratic nonlinear coupling between the fundamental and the second-harmonic fields, with $\xi = \Delta k L/2$, where $L$ is the length of the nonlinear medium in the cavity and $\Delta k = 2k(\omega_0)-k(2\omega_0)$ the wave-vector mismatch of the SHG process $\omega_0+\omega_0 = 2\omega_0$. The third term in Eq.~\eqref{MF2} describes temporal walk-off between the fundamental and the second-harmonic fields ($d$ is a normalised walk-off parameter), that originates from the different group velocities at those frequencies (note that $\tau$ is a delayed time, defined in a reference frame moving at the group velocity of light at the fundamental frequency). The last term in Eq.~\eqref{MF1} accounts for the external coherent driving of the cavity at the fundamental frequency. Following the notation in~\cite{leo_walkoff_2015} (see also Appendix~\ref{Apx1}), the normalisation is such that $t\rightarrow \alpha_1 t/t_\mathrm{R}$, $\tau \rightarrow \tau[2\alpha_1/(|{k}_1''|L)]^{1/2}$, $v_1 = A\kappa L/\alpha_1$, $v_2 = B\kappa L/\alpha_1$, $\Delta_{1,2} = \delta_{1,2}/\alpha_1$, $d = \Delta k'\left[2L/(\alpha_1|k_1''|)\right]^{1/2}$, $\eta_1 = \mathrm{sgn}[k_1'']$, $\eta_2 = k_2''/|k_1''|$, $\alpha = \alpha_2/\alpha_1$, $S = \sqrt{\theta_1}A_\mathrm{in}\kappa L/\alpha_1^2$, $\tau_\mathrm{s} = t_\mathrm{R}[2\alpha_1/(|{k}_1''|L)]^{1/2}$. Here $t_\mathrm{R} = \mathrm{FSR}^{-1}$ is the cavity round-trip
time for pulses at $\omega_0$, with $\mathrm{FSR}$ the corresponding free spectral range; $\alpha_{1,2}$ describe half of the total cavity power losses per round trip for light at $\omega_0$ and $2\omega_0$, respectively; ${k}''_{1,2} = \mathrm{d}^2k/\mathrm{d}\omega^2|_{\omega_0, 2\omega_0}$ are the GVD coefficients; $\Delta {k}' = \mathrm{d}k/\mathrm{d}\omega|_{2\omega_0}-\mathrm{d}k/\mathrm{d}\omega|_{\omega_0}$ is the group-velocity mismatch; $\kappa$ is an effective second-order nonlinearity coefficient, normalised such that the dimensional fields $|A|^2$, $|B|^2$, and $|A_\mathrm{in}|^2$ are measured in watts~\cite{leo_walkoff_2015}; $\delta_{1,2}$ are the phase detunings of carrier waves at $\omega_0$ and $2\omega_0$ from the closest linear cavity resonances; and $\theta_1$ is the power transmission coefficient of the coupler used to inject the driving field $A_\mathrm{in}$ into the cavity.

Equations~\eqref{MF1} and~\eqref{MF2} can be numerically propagated over the slow time $t$ with a split-step Fourier type method that uses, e.g., a fourth-order Runge Kutta scheme to evaluate the nonlinear step (results that follow were obtained using that scheme). Note that, in addition to the variables explicitly contained in Eqs.~\eqref{MF1} and ~\eqref{MF2}, the cavity dynamics also depends on the (normalized) cavity round-trip time $\tau_\mathrm{s}$, as it does in Kerr microresonators~\cite{hansson_dynamics_2013, coen_modeling_2013, torres-company_comparative_2014}. The round-trip time enters the model described by Eqs.~\eqref{MF1} and~\eqref{MF2} by imposing periodic boundaries on the fast time variable, such that $\tau\in[-\tau_\mathrm{s}/2, \tau_\mathrm{s}/2]$. In this context, the $\tau_\mathrm{s}^{-1}$ spacing of the normalised Fourier frequency grid represents a single FSR. At this point we emphasise that, although Eqs.~\eqref{MF1} and~\eqref{MF2} account for the dominant physical processes at play in dispersive, doubly resonant cavity SHG, additional perturbative terms may be necessary to ensure the model's validity under specific conditions. Rigorous analysis of spectrally broadband signals can, for example, require the inclusion of higher-order dispersion terms ($\propto \partial^n/\partial\tau^n$). We also note that Eqs.~\eqref{MF1} and \eqref{MF2} can lose validity when the combs around the fundamental and the SH frequencies start to overlap, or when other nonlinear processes such as third-harmonic generation, Kerr nonlinearity or down-conversion to around $\omega_0/2$ compete with SHG. In these situations, it may be necessary to use a more complex approach based on a generalised nonlinear envelope equation~\cite{conforti_nonlinear_2010_1, conforti_nonlinear_2010_2, hansson_SEE, wabnitz_harmonic_2010}.

\section{Modulation instability analysis}
\label{MIsec}
We are interested in the formation of optical frequency combs and corresponding temporal pulse trains in doubly resonant SHG cavities as described by Eqs.~\eqref{MF1} and ~\eqref{MF2}. Similarly to Kerr resonators~\cite{coen_modeling_2013} and singly resonant SHG configurations~\cite{leo_walkoff_2015}, MI of the cavity cw steady state is a key requirement for frequency comb emergence. Physically, MI in cavity-enhanced SHG arises from the down-conversion of the second-harmonic field, an effect which has been extensively investigated in the context of internally-pumped optical parametric oscillators~\cite{schiller_quadruply_1993, marte_competing_1994, lodahl_spatiotemporal_2001}. In \cite{trillo_pulse-train_1996}, Trillo and Haelterman studied equations formally identical to Eqs.~\eqref{MF1} and \eqref{MF2}, identifying the presence of temporal MI for the case of zero group-velocity mismatch ($d = 0$). Analogous results (using formally similar equations with $d=0$) were later reported in the context of spatially diffractive cavities~\cite{etrich_pattern_1997}. However, whilst negligible spatial walk-off may be an appropriate approximation in the context of diffractive pattern formation, for most dispersive systems one realistically finds that the temporal walk-off $|d|\gg1$. For example, parameters from the singly resonant configuration in ~\cite{ricciardi_frequency_2015, leo_walkoff_2015} would yield $d\approx 2000$. At first glimpse, such a large walk-off value could be envisioned to prevent the applicability of the mean-field approach altogether. However, we have carefully compared results derived from Eqs.~\eqref{MF1} and~\eqref{MF2} with those obtained from the full infinite-dimensional cavity map, and we have systematically found the two approaches to be in excellent agreement. As we shall see below, this agreement likely stems from the nonlinearity cancelling the effects of linear walk-off, thus resulting in slow field evolution that can be accurately captured by the mean-field equations.

To investigate MI in the presence of a nonzero walk-off, we examine the stability of the cavity's mixed-mode cw stationary solutions $v_{10}$ and $v_{20}$ for a wide range of walk-off parameters. The mixed-mode solution can easily be found by setting all the derivatives in Eqs.~\eqref{MF1} and ~\eqref{MF2} to zero~\cite{trillo_pulse-train_1996, etrich_pattern_1997,etrich_solitary_1997}, yielding
\begin{align}
&v_{10} = \frac{(\alpha + {i}\Delta_2)S}{(\alpha + {i}\Delta_2)(1 + {i}\Delta_1) + \mathrm{sinc}^2(\xi)Y_1} \label{FFmode}\\
&v_{20} = \frac{{i}\rme^{{i}\xi}\mathrm{sinc}(\xi)v_{10}^2}{(\alpha + {i}\Delta_2)},
\end{align}
where $Y_1 = |v_{10}|^2$ is a root of the cubic equation
\begin{multline}\label{CWmodes}
\mathrm{sinc}^4(\xi)Y_1^3 + 2(\alpha-\Delta_1\Delta_2)\mathrm{sinc}^2(\xi)Y_1^2\\
+(1+\Delta_1^2)(\alpha^2+\Delta_2^2)Y_1 = (\alpha^2+\Delta_2^2)|S|^2.
\end{multline}
We then analyse the linear stability of the cw modes with respect to modulations at frequency $\Omega$ by introducing the ansatz $v_k = v_{k0} + a_k\exp(\lambda t+i\Omega \tau) + b_k\exp(\lambda^* t-i\Omega \tau)$ with $k = 1,2$ into Eqs.~\eqref{MF1} and ~\eqref{MF2}. By linearising with respect to $a_k$ and $b_k$, we obtain a $4\times4$ matrix equation whose eigenvalues obey the quartic equation
\begin{equation}\label{eigenvalues}
\lambda^4 +c_3(\Omega)\lambda^3 + c_2(\Omega)\lambda^2 + c_1(\Omega)\lambda + c_0(\Omega) = 0.
\end{equation}
The coefficients $c_k$ with $k = 0, 1, 2, 3$ depend on the perturbation frequency $\Omega$ and the pump-resonator parameters as follows
\begin{align}
&c_0(\Omega) = 4I_1(I_1 + \overline{\alpha} - \overline{\Delta}_1\overline{\Delta}_2) + (1+\overline{\Delta}_1^2 - I_2) (\overline{\alpha}^2 + \overline{\Delta}_2^2),\nonumber\\
&c_1(\Omega) = 4I_1(1+\overline{\alpha}) +2\overline{\alpha}(1+\overline{\Delta}_1^2+\overline{\alpha}) - 2I_2\overline{\alpha},\nonumber\\
&c_2(\Omega) = 4I_1-I_2+1+\overline{\Delta}_1^2+\overline{\Delta}_2^2+\overline{\alpha}(4+\overline{\alpha}), \nonumber \\
&c_3(\Omega) = 2(1+\overline{\alpha}), \label{cs}
\end{align}
where $\overline{\alpha} = \alpha + id\hspace{1pt}\Omega$, $\overline{\Delta}_1 = {\Delta}_1 -\eta_1\Omega^2$, $\overline{\Delta}_2 = {\Delta}_2 -\eta_2\Omega^2$, $I_1=\mathrm{sinc}^2(\xi)|v_{10}|^2$ and $I_2=\mathrm{sinc}^2(\xi)|v_{20}|^2$.

\begin{figure}[htb]
 \centering
 \includegraphics[width = 8.3cm, clip=true]{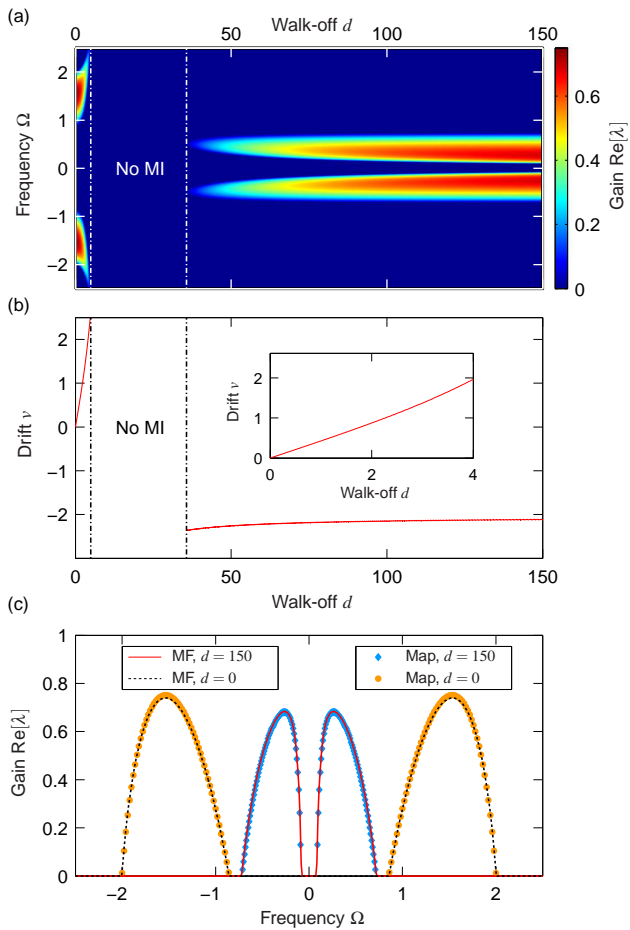}
 \caption{(a) MI gain for a doubly resonant SHG system as a function of frequency and walk-off, with $\eta_1 = -1$, $\eta_2 = -0.5$, $\alpha = 0.5$, $\xi=0$, $\Delta_1 = 2$, $\Delta_2 = 4$ and $S = 5$. (b) Predicted drift velocity of an emerging pattern at the most unstable frequency. Inset shows zoom for small walk-off. (c) MI gain profiles obtained from the stability analysis of the mean-field (MF) equations (solid and dashed curves) and the full cavity map (filled circles and diamonds) for two different walk-off values as indicated.}
 \label{FigMI}
\end{figure}

By taking into account nonzero walk-off, Eqs.~\eqref{eigenvalues} and~\eqref{cs} generalize stability analyses presented in previous studies~\cite{etrich_pattern_1997, etrich_solitary_1997, trillo_pulse-train_1996, lodahl_thesis}. In general, the MI gain spectrum, i.e. $\mathrm{Re}[\lambda(\Omega)]$, predicted by Eq.~\eqref{eigenvalues} exhibits complex dependence on the cavity detuning parameters $\Delta_{1,2}$ as well as on the phase-mismatch parameter $\xi$. Because we are first and foremost interested in the role of walk-off, we assume a resonator that is made entirely out of the nonlinear material (such as a monolithic microresonator), and assume the SHG to be naturally phase-matched. In such a configuration, $\Delta k = \xi = 0$, and $\Delta_2 = 2\Delta_1$ (see Appendix~\ref{Apx2}). A typical example of the predicted gain as a function of frequency $\Omega$ and walk-off $d$ is shown in Fig.~\ref{FigMI}(a). To facilitate comparison with earlier studies of spatial cavities where walk-off was not considered ~\cite{etrich_pattern_1997,etrich_solitary_1997}, we have adopted a similar set of parameters (listed in the caption; note that with these parameters the cw response is monostable). In agreement with those studies (but in contrast to the case of singly resonant SHG~\cite{leo_walkoff_2015}), we see that MI can arise even when $d = 0$. As the walk-off $d$ increases, however, the gain quickly reduces and eventually disappears altogether: the cw mode becomes modulationally stable. Surprisingly, for sufficiently large walk-off values ($|d|\gg 1$) MI re-emerges, but at a very different spectral position [see also Fig.~\ref{FigMI}(c)]. It thus appears that, in addition to MI identified previously with $d = 0$~\cite{etrich_pattern_1997, etrich_solitary_1997, trillo_pulse-train_1996}, the cavity also supports another MI regime that manifests itself for $|d|\gg1$. At this point, we emphasize that we have confirmed our mean-field-based MI analysis by comparing the results against those obtained from the full infinite-dimensional cavity map. Specifically, we derived the MI gain spectrum from the full cavity map by using an approach similar to that previously employed for Kerr resonators~\cite{mclaughlin_new_1985,coen_modulational_1997, hansson_frequency_2015}, and we found the results to be in almost exact agreement with those shown in Fig.~\ref{FigMI}. An example of the agreement is explicitly shown in Fig.~\ref{FigMI}(c), where we compare the MI gain spectra obtained from Eq.~\eqref{eigenvalues} (solid and dashed curves) and from the full cavity map (circles and diamonds) for two different walk-off values ($d = 0$ and $d = 150$). (For full details on the stability analysis of the full map, see Appendix~\ref{Apx3}.) The excellent agreement clearly validates the use of a mean-field model, even when $|d|\gg1$.

MI leads to the amplification of spectral sidebands around $\omega_0$ and $2\omega_0$, which in the time domain manifests itself as the growth of periodic modulations on top of the cw fields. Similarly to the singly resonant case~\cite{leo_walkoff_2015}, the generated patterns are found to exhibit a temporal drift whose magnitude is much smaller than the walk-off $d$. As a matter of fact, the drift rate $v = \mathrm{d}\tau_\mathrm{c}/\mathrm{d}t$, where $\tau_\mathrm{c}(t)$ represents the fast time coordinate of a particular point along the sinusoidal perturbation emerging from MI, can be obtained from the imaginary part of the corresponding eigenvalue as $v=\mathrm{Im}[\lambda]/\Omega$, with $\Omega$ being the frequency of the pattern~\cite{zambrini_convection-induced_2005}. In Fig.~\ref{FigMI}(b) we show the drift rate $v$ predicted at frequencies experiencing the maximum MI gain as a function of walk-off. For small $d$, the drift rate increases monotonously as $v\approx d/2$ (see inset) up until the point where MI disappears ($d\approx 4$). But interestingly, for walk-off values associated with the re-emergence of MI ($d>35$), the sign of the drift changes and its magnitude remains almost constant, with $|v|\approx 2 \ll |d|$. This means that the nonlinearity compensates for the linear spectral phase induced by the walk-off, which is likely to explain why we systematically find that, even for very large $|d|$, the mean-field approach yields results that are in almost identical agreement with those obtained from the full map (see Appendix~\ref{Apx1} for the full map equations). Indeed, the nonlinear cancellation of the linear walk-off ensures that the key condition behind the mean-field approximation is satisfied: the fields only evolve slowly, with no significant changes over a single round trip.

\section{Frequency comb simulations}
\label{sims}
After the initial stage of exponential sideband amplification, the modulation triggered by MI may reshape into complex temporal patterns and corresponding frequency combs. In fact, extensive simulations of the mean-field Eqs.~\eqref{MF1} and ~\eqref{MF2} reveal several distinct dynamical regimes with a rich nonlinear behaviour. To give an example of the diversity, we carried out a single long simulation that mimics the experimental procedure of sweeping the driving laser across a cavity resonance~\cite{herr_temporal_2014, lamont_route_2013, luo_spontaneous_2015}. Specifically, we use an initial condition consisting of white noise (amplitude $10^{-6}$), start the simulation far from resonance with $\Delta_1 = -3.5$ (natural phase-matching is assumed throughout, so that $\xi = 0$ and $\Delta_2 = 2\Delta_1$), and we then slowly increase the detuning up to $\Delta_1 = 3.5$ over a (normalised) slow time scale of $t = 500000$ (the precise dynamics may depend on the speed of the detuning ramp). In this simulation, we have set $d = 450$ and also assumed the group-velocity dispersion at the SH wavelength to be normal, $\eta_2 = 2$, which is a more realistic situation for temporal cavity dynamics (recall that the parameters in Fig.~\ref{FigMI} were chosen so as to facilitate comparison with prior studies in spatial cavity systems). At this point we may remark, however, that for $|d|\gg1$ we find the group velocity dispersion $\eta_2$ to have negligible impact. This can be understood by noting that the dispersion at the second-harmonic reads $d\hspace{1pt}\Omega+\eta_2\Omega^2$, with $\Omega$ and $\eta_2$ of the order of unity by normalization.

\begin{figure}[htb]
  \centering
 \includegraphics[width = 8.3cm]{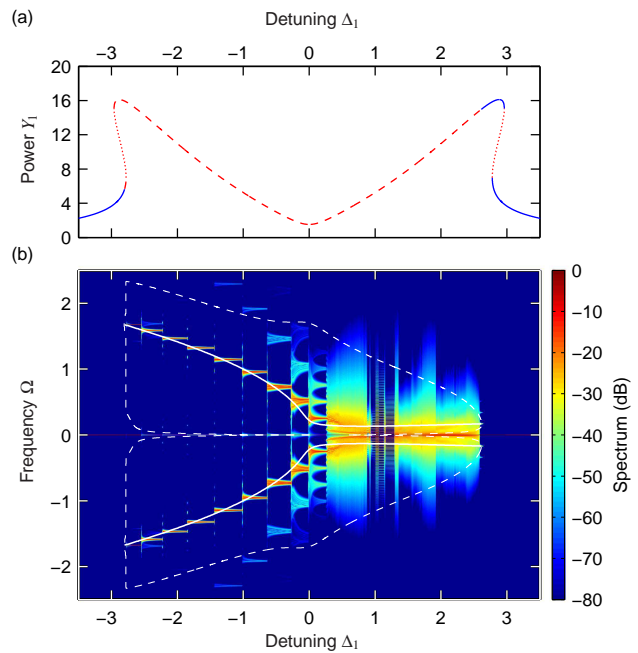}
   \caption{(a) Stable and unstable cw modes of the system for different detunings. The parameters are $d=450$, $\eta_1 = -1$, $\eta_2 = 2$, $\alpha = 0.5$, $\xi=0$,  and $S = 5$. (b) Spectral dynamics as the cavity detuning $\Delta_1$ is swept across a resonance (note that $\Delta_2 = 2\Delta_1$ throughout the sweep). The width of the simulation time window $\tau_\mathrm{s} = 1000$ implies an FSR of $\Omega_\mathrm{FSR}=2\pi/\tau_\mathrm{s} \approx 0.0063$.}
  \label{Figsweep}
\end{figure}

We first show, in Fig.~\ref{Figsweep}(a), the cw cavity response [i.e., $Y_1$ given by Eq.~\eqref{CWmodes}] for the different detunings accessed during the simulation. Because $\Delta_2 = 2\Delta_1$, the cw resonance displays a complex double-peaked shape that is symmetric around $\Delta_1 = 0$. (It is worth noting that for constant $\Delta_2 \neq 0$, the resonance would exhibit a tilt~\cite{etrich_pattern_1997} similar to the case of Kerr cavities~\cite{coen_universal_2013}.) The cw modes are stable (blue solid line) for large $|\Delta_1|$, but also for detunings close to the peak on the positive detuning side. Portions of the resonance with a negative slope are unstable against cw perturbations (dotted red line), whilst the cw modes are modulationally unstable (dashed red line) dominantly for detunings in between the two resonance maxima.

Figure~\ref{Figsweep}(b) shows the simulated intracavity spectrum around the fundamental frequency as the detuning is slowly increased. For each detuning, we also highlight the regions that exhibit MI (areas within the dashed white boxes), as well as the critical point where the gain is the largest (solid white line). In agreement with the linear stability analysis, the cw solution becomes modulationally unstable around ${\Delta_1 \approx -2.8}$, at which point spectral sidebands grow at the frequency of maximum gain that is separated by multiple FSRs from the pump. (The normalized round-trip time is $\tau_\mathrm{s} = 1000$, corresponding to an FSR of $\Omega_\mathrm{FSR}=2\pi/\tau_\mathrm{s} \approx 0.0063$.) As $\Delta_1$ increases, the sidebands are seen to experience several abrupt frequency shifts towards the pump, with the sidebands' frequency detunings following the maximum MI gain frequency (solid white line). In the time-domain, each abrupt frequency shift manifests itself as a reduction of the repetition rate of the temporal MI pattern. A detailed investigation is beyond the scope of our work, yet we note that this frequency transition is reminiscent of the so-called Eckhaus instability~\cite{eckhaus_studies_1965, lowe_pattern_1985}. In that process, a periodic pattern whose frequency is far from the critical point goes unstable, which leads to the elimination of that pattern and to the eventual formation of a new pattern closer to the critical point~\cite{lowe_pattern_1985}. This is indeed similar to what occurs in the simulation shown in Fig.~\ref{Figsweep}(b). Because the maximum MI gain frequency evolves with $\Delta_1$, a given sideband will eventually be so detuned from the critical frequency that the corresponding pattern goes unstable, thus resulting in the emergence of another pattern with frequency closer to the maximum MI gain.

\begin{figure}[h]
 \centering
  \includegraphics[width = 8.3cm]{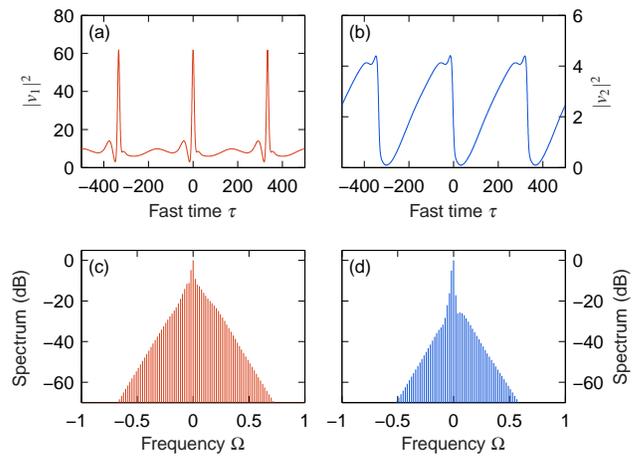}
 \caption{Temporal patterns in a doubly resonant SHG system with large walk-off. (a, b) Temporal profiles at the (a) fundamental and (b) second-harmonic wavelengths. Corresponding spectra are shown in (c) and (d), respectively.}
 \label{Figpattern}
\end{figure}

Around $\Delta_1\approx 0.25$, Fig.~\ref{Figsweep}(b) shows that the system transitions to a very different regime, where the spectrum is dominantly composed of spectral lines spaced by a single FSR. This transition is in fact similar to the one observed in Kerr microresonators~\cite{herr_universal_2012}, when the field evolves from stable to chaotic MI patterns~\cite{coen_universal_2013, erkintalo_coherence_2014}. Analogously, closer examination of results in Fig.~\ref{Figsweep}(b) shows that, for $\Delta_1>0.25$, the intracavity field mostly consists of unstable temporal patterns that exhibit strong fluctuations. Yet, narrow regions of stability can also be identified. For example, for $\Delta_1 \approx 1.2$ we find stable frequency combs whose lines are spaced by multiple FSRs. In Fig.~\ref{Figpattern} we show temporal and spectral characteristics of this regime in more details. These results were obtained by running another simulation, where the detuning sweep was stopped at $\Delta_1 = 1.2$. It can be seen that the time-domain intracavity fields are composed of three, periodically-spaced temporal structures [Fig.~\ref{Figpattern}(a, b)], corresponding to a frequency comb with $3\times\mathrm{FSR}$ spacing [Fig.~\ref{Figpattern}(c, d)]. Simulations carried out over very long time scales reveal the combs and the patterns to be fully stable and coherent. Yet, similarly to the singly resonant case~\cite{leo_walkoff_2015}, the patterns exhibit a temporal drift whose magnitude is much smaller than the walk-off parameter $|d|$, in accordance with Fig.~\ref{FigMI}(b).

\begin{figure}[htb]
 \centering
  \includegraphics[width = 8.3cm,clip=true]{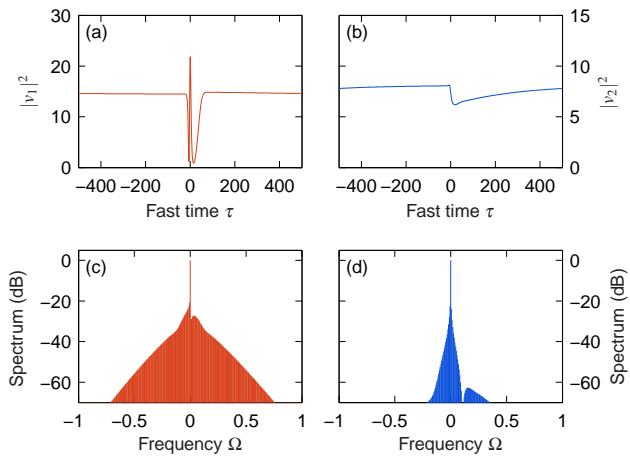}
 \caption{Localised wave in a doubly resonant SHG system with large walk-off. (a, b) Temporal profiles at the (a) fundamental and (b) second-harmonic wavelengths. Corresponding spectra are shown in (c) and (d), respectively.}
 \label{Figsol}
\end{figure}

Interestingly, around $\Delta_1 \approx 2.5$ we have also found stable frequency combs whose lines are spaced by a single FSR. This suggests the existence of localised pulses --- dissipative solitary waves --- in the time domain. Figure~\ref{Figsol} shows the temporal (a, b) and spectral (c, d) characteristics of the dissipative temporal structure that exists for $\Delta_1 = 2.53$. At the fundamental frequency [Fig.~\ref{Figsol}(a, c)], the structure consists of a complex temporal feature, localised to within $\tau \approx 50$ of the $\tau_\mathrm{s} = 1000$ round-trip time, embedded on a cw background; at the second-harmonic, only a small step perturbs the cw [Fig.~\ref{Figsol}(b, d)]. Aside from a small temporal drift, also these fields are fully stable. Indeed, we have used long split-step simulations up to $t = 2\times10^5$ as well as a more rigorous Newton-Raphson technique~\cite{coen_modeling_2013} to verify that, with a trivial change of reference frame that cancels their drift~\cite{jang_ultraweak_2013}, the fields shown in Fig.~\ref{Figpattern} and Fig.~\ref{Figsol} correspond to stable steady-state solutions of Eqs.~\eqref{MF1} and ~\eqref{MF2}. Finally, we have confirmed that the solutions are not artifacts of the mean-field model: almost identical structures can be observed in simulations of the full cavity map.

Despite its stability and high-degree of localisation, the temporal structure shown in Fig.~\ref{Figsol}(a, b) does not correspond to an isolated soliton, akin to those manifesting themselves in, e.g., Kerr resonators~\cite{herr_temporal_2014, leo_temporal_2010,leo_dynamics_2013, jang_ultraweak_2013}. Indeed, in contrast to such \emph{cavity solitons}, the structure predicted here depends sensitively on the boundaries, i.e. the cavity round-trip time $\tau_\mathrm{s}$. This can be qualitatively understood by noting that the SH field profile in Fig. ~\ref{Figsol}(b) is not localized, but it instead exhibits variations across the whole fast time domain. To ensure that the physically-enforced periodic boundary conditions remain fulfilled, any change in $\tau_\mathrm{s}$ must be met with a change in the SH profile, hence necessitating a different fundamental field. Although we find that reducing the round-trip time to, e.g., $\tau_\mathrm{s} = 500$ yields a localized solution only somewhat different from that shown in Fig.~\ref{Figsol}, extending the boundaries  to $\tau_\mathrm{s} = 2000$ surprisingly gives rise to a periodic pattern consisting of multiple temporal structures. This suggests that the solution in Fig.~\ref{Figsol} corresponds to a single period of a pattern rather than an isolated soliton. Regardless, the numerical simulation results presented in this section clearly illustrate the feasibility of obtaining coherent optical frequency combs and corresponding temporal structures in doubly resonant SHG.

\section{Reduced mean-field approximation}
\label{Reduced}
The previous results were obtained using the coupled mean-field system described by Eqs.~\eqref{MF1} and~\eqref{MF2}. Interestingly, we have also found that, for a wide excursion of parameters, Eqs.~\eqref{MF1} and ~\eqref{MF2} can be further reduced into a single mean-field equation. This simplification is enabled by the observation that, similarly as for highly phase-mismatched single-pass SHG systems~\cite{nikolov_quadratic_2003},  the second-harmonic field $v_2(t,\tau)$ varies slowly with $t$, so that one can make the approximation that the term $\partial v_2/\partial t$ can be neglected in Eq.~\eqref{MF2}. Under this assumption, the second-harmonic field is dynamically slaved to the fundamental field, and it can be explicitly solved from Eq.~\eqref{MF2}:
\begin{align}
v_2(t,\tau) = ie^{i\xi}\mathrm{sinc}(\xi)\left[v_1^2(t,\tau)\otimes J(\tau)\right], \label{slaved}
\end{align}
\noindent where $\otimes$ denotes convolution and $J(\tau)=\mathscr{F}^{-1}[\hat{J}(\Omega)]$ is the inverse Fourier transform of the frequency-domain response function $\hat{J}(\Omega)$, defined as
\begin{equation}
\hat{J}(\Omega) = \frac{1}{\alpha +i\Delta_2 -id\hspace{1pt}\Omega-i\eta_2\Omega^2}.
\label{response}
\end{equation}
Substituting Eq.~\eqref{slaved} into Eq.~\eqref{MF1}, we obtain:
\begin{equation}
\frac{\partial v_1}{\partial t} = \left[-1 - {i}\Delta_1 -{i}\eta_1\frac{\partial^2}{\partial \tau^2}\right]v_1-\rho v_1^*\left[v_1^2\otimes J\right] +S,\label{MF}
\end{equation}
where $\rho = \mathrm{sinc}^2(\xi)$. Equation ~\eqref{MF} is formally equivalent to the mean-field equation derived in~\cite{leo_walkoff_2015} for the description of \emph{singly resonant} cavity SHG; it is only the nonlinear response function [here $J(\tau)$] that is different between the two equations.  A detailed comparison of the response functions is beyond the scope of our present article, yet we note that they are substantially different, reflecting the distinct behaviour of singly- and doubly resonant cavities.  For example, in contrast with the singly resonant case, the doubly resonant response function may give rise to instantaneous Kerr-like phase shifts even if the SHG process is phase-matched, provided that $\Delta_2\neq 0$. As a matter of fact, this feature is illustrated in the tilted resonances shown in Fig.~\ref{Figsweep}; under phase-matching, no such tilts occur in the singly resonant case. It is also interesting to note that, in the singly resonant case, the nonlinearity tends towards a pure Kerr-like response in the limit of large phase-mismatch ($\xi\gg 1$), whereas the  doubly resonant response function does not depend on $\xi$ at all [see Eq.~\eqref{response}]. Notwithstanding, it can be seen from Eq.~\eqref{response} that pure Kerr-like nonlinearity can ensue for very large second-harmonic detunings, when $|\Delta_2| \gg d\hspace{1pt}\Omega + \eta_2\Omega^2$ and $|\Delta_2|\gg \alpha$. In this limit, Eq.~\eqref{MF} tends towards the AC-driven nonlinear Schr\"odinger equation (also known as the Lugiato-Lefever equation), which has been widely applied to the study of Kerr nonlinear cavities~\cite{coen_modeling_2013,haelterman_dissipative_1992, lugiato_spatial_1987}.

In spite of the approximation behind the mean-field Eq.~\eqref{MF}, we have found through extensive simulations that, particularly for $|d|\gg1$, Eqs.~\eqref{response} and~\eqref{MF} are fully adequate for the modelling of \emph{doubly resonant} cavity SHG, and they deliver predictions that are in almost identical agreement with those obtained from the more accurate system given by Eqs.~\eqref{MF1} and~\eqref{MF2}. As an example, all the simulation results presented in Section~\ref{sims} can be almost exactly reproduced by using Eq.~\eqref{MF}. Notably, although strictly speaking $\partial v_2/\partial t = 0$ only for non-drifting steady-state solutions, Eq.~\eqref{MF} reproduces the drift of the solutions and even captures the initial dynamics with excellent accuracy.

\begin{figure}[htb]
 \centering
  \includegraphics[width = 8.3cm, clip=true]{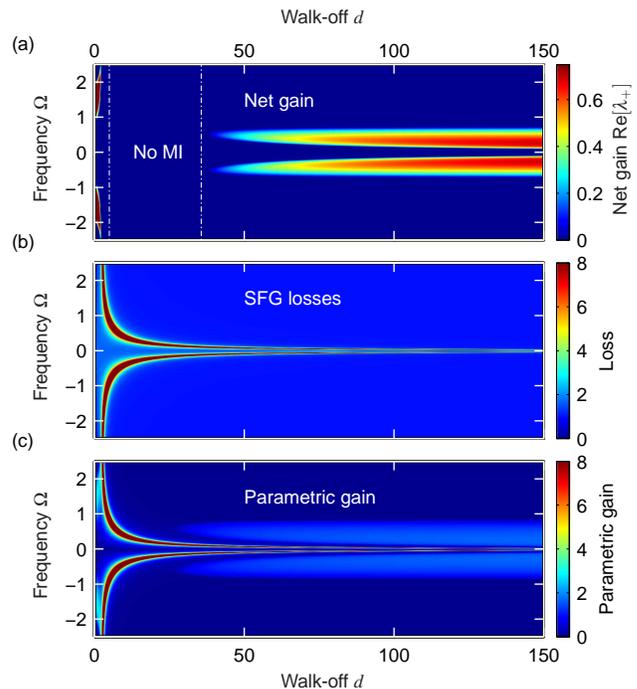}
 \caption{(a) MI gain evaluated from the simplified eigenvalue expression Eq.~\eqref{MIgain} using parameters identical to those in Fig.~\ref{FigMI}. Competing total losses and sideband gain are shown in (b) and (c), respectively. Note that the MI gain in (a) can be obtained by subtracting the data in (b) from the data in (c). For reference, the dash-dotted white lines in (a) highlight the same walk-off values as those shown in Fig.~\ref{FigMI}(a). }
 \label{simple_MI}
\end{figure}

In addition to simplifying numerical integration implementations, Eq.~\eqref{MF} also permits a more straightforward stability analysis of the intracavity fields. Indeed, following the linear stability analysis in ~\cite{leo_walkoff_2015}, the potentially unstable eigenvalues that define the MI gain are found to read
\begin{widetext}
\begin{equation}
\lambda_\pm(\Omega) = -\left(1+\rho Y_1[\hat{J}(\Omega)+\hat{J}^*(-\Omega)]\right) \pm \sqrt{|\hat{J}(0)|^2\rho^2Y_1^2-\left(\Delta_1-\eta_1\Omega^2-i\rho Y_1[\hat{J}(\Omega)-\hat{J}^*(-\Omega)]\right)^2}.
\label{MIgain}
\end{equation}
\end{widetext}
These eigenvalues can be directly compared with those derived from the stability analysis of the coupled mean-field equations [see Eq.~\eqref{eigenvalues}]. To this end, in Fig.~\ref{simple_MI}(a) we show the MI gain predicted by Eq.~\eqref{MIgain}, i.e. $\mathrm{Re}[\lambda_+(\Omega)]$, for the same parameters that were used in obtaining Fig.~\ref{FigMI}(a). Besides overestimation of MI gain for small walk-off parameters $d$, the gain predicted by Eq.~\eqref{MIgain} shows excellent agreement with that obtained from the more general Eqs.~\eqref{MF1} and~\eqref{MF2} [see Fig.~\ref{FigMI}(a)].

Moreover, the reduced eigenvalue Eq.~\eqref{MIgain} enables a more straightforward assessment of the physics that underpins MI. In a manner similar to the case of singly resonant cavity SHG~\cite{leo_walkoff_2015}, the real part of the square root term describes the second-harmonic-pumped optical parametric oscillator gain experienced by sidebands at $\omega_0\pm\Omega$, whilst the terms outside the square root describe linear and nonlinear losses, with the latter arising from competing sum-frequency generation (SFG) processes of the form $(\omega_0+\Omega)+\omega_ 0 = 2\omega_0 + \Omega$~\cite{ricciardi_frequency_2015, leo_walkoff_2015}. In Figs.~\ref{simple_MI}(b) and (c) we plot the total sideband loss ($1+\rho Y_1\mathrm{Re}[\hat{J}(\Omega)+\hat{J}^*(-\Omega)]$) and parametric gain [real part of the square root term in Eq.~\eqref{MIgain}] profiles, respectively, for a range of walk-off parameters $d$. [Note that the net gain shown in Fig.~\ref{simple_MI}(a) can be obtained simply by subtracting the losses shown in Fig.~\ref{simple_MI}(b) from the parametric gain in Fig.~\ref{simple_MI}(c).]

\begin{figure}[htb]
 \centering
  \includegraphics[width = 8.3cm, clip=true]{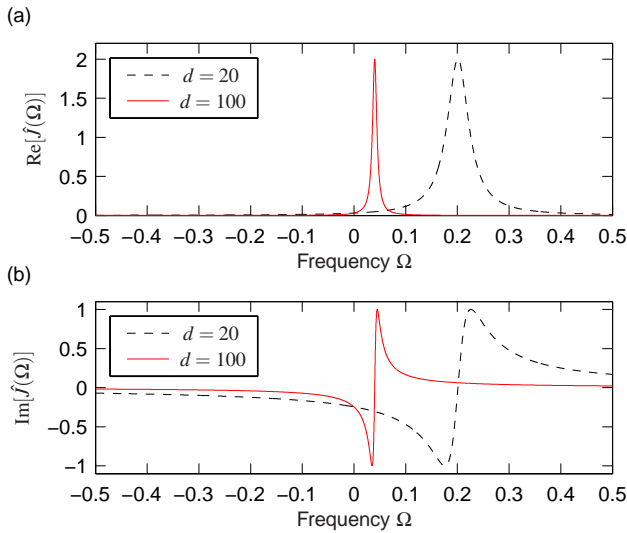}
 \caption{(a) Real and (b) imaginary parts of $\hat{J}(\Omega)$ for $d = 20$ (dashed black curve) and $d = 100$ (solid red curve) as a function of frequency. Other parameters are $\Delta_2 = 4$ and $\alpha = 0.5$. Note that the Lorentzian profiles are centred at $\Omega = 0.2$ (dashed black curve) and $\Omega = 0.04$ (solid red curve), in accordance with the SH resonance condition $\Omega_\mathrm{R}\approx \Delta_2/d$.}
 \label{responseFig}
\end{figure}

Several observations can be made. First, similarly with the singly resonant configuration~\cite{leo_walkoff_2015}, nonlinear losses are significantly impacted by walk-off [Fig.~\ref{simple_MI}(b)]. But in contrast with that configuration, where the loss profile as a function of frequency was found to follow a simple $\mathrm{sinc}^2$ shape linked to the SFG efficiency, we find that in the doubly resonant scheme it is the condition of resonance around the second harmonic that mostly defines the loss profile. Specifically, a sum-frequency component at $2\omega_0+\Omega_\mathrm{R}$ is on resonance when $\Delta_2-d\hspace{1pt}\Omega_\mathrm{R}-\eta_2\Omega^2_\mathrm{R} = 0$, leading to enhanced SFG losses around $\omega_0+\Omega_\mathrm{R}$. Closer inspection of data shown in Fig.~\ref{simple_MI}(b) indeed reveals that (i) the loss bands have Lorentzian resonance profiles and (ii) one of them is always centered at $\Omega_\mathrm{R}$, with the second band at $\approx-\Omega_\mathrm{R}$ arising due to the coupling between sidebands at $\omega_0\pm\Omega$ [see $\hat{J}^*(-\Omega)$ terms in Eq.~\eqref{MIgain}]. In this context, it is worth pointing out that, because the SFG losses are proportional to $\mathrm{Re}[\hat{J}(\Omega)+\hat{J}^*(-\Omega)] = \mathrm{Re}[\hat{J}(\Omega)] + \mathrm{Re}[\hat{J}(-\Omega)]$, the loss profile is simply governed by the real part of $\hat{J}(\Omega)$ [defined in Eq.~\eqref{response}]. For $|d|\gg 1$, that profile assumes a Lorentzian shape [see also Fig.~\ref{responseFig}(a)].

In addition to affecting losses, walk-off also has a significant impact on the parametric gain profile [Figure~\ref{simple_MI}(c)], which is in stark contrast with the singly resonant configuration~\cite{leo_walkoff_2015}. We suspect this is because, in the doubly resonant scheme, the SFG process imparts a nonlinear phase shift for sidebands around $\omega_0\pm\Omega_\mathrm{R}$, resulting in the disruption of the usual parametric phase-matching. This nonlinear phase shift is manifested in the last term inside the square root in Eq.~\eqref{MIgain}. In contrast to the nonlinear losses, the square root term describing parametric gain is found to depend both on the real and the imaginary parts of $\hat{J}(\Omega)$. We also remark that the profile of the imaginary part of $\hat{J}(\Omega)$ is reminiscent of the real part of a resonant atomic susceptibility, as shown in Fig.~\ref{responseFig}(b), which is not particularly surprising given the resemblance between $\hat{J}(\Omega)$ and the frequency response of a damped and driven (with an impulse at $t=0$) harmonic oscillator.

The combined effect of the SFG-induced losses and phase shifts is to hinder the oscillation of the parametric sidebands, such that net MI gain only arises at frequencies far away from those associated with resonant SFG. This notion allows us to better understand the role of walk-off in facilitating MI. Specifically, Eq.~\eqref{response} shows that the resonance is centred at $\Omega_\mathrm{R} \approx \Delta_2/d$ and has a linewidth of $w\approx \alpha/d$. It is thus clear that for increasing values of the walk-off parameter $d$, the resonance bands shift closer to the pump and decrease in width, thereby affecting a narrowing sliver of frequencies around $\Omega = 0$. We illustrate this behaviour in Figs.~\ref{responseFig}(a) and (b), where we respectively show the real and imaginary parts of $\hat{J}(\Omega)$ for two different walk-off parameters [other parameters as in Fig.~\ref{simple_MI}]. Parametric gain capable of driving MI is obtained when the resonance-associated losses and phase shifts no longer affect frequency components that would exhibit gain in its absence [i.e., when neglecting the last term inside the square root in Eq.~\eqref{MIgain}]. In Fig.~\ref{simple_MI}(c) we indeed see how, for sufficiently large walk-off, gain bands emerge beyond the narrow Lorentzian resonance features.

Of course, we emphasise that the precise gain features discussed above [and illustrated in Figs.~\ref{FigMI} and ~\ref{simple_MI}] depend on the particular parameters involved. Nevertheless, we have found that our general insights are capable of elucidating the MI characteristics over a wide range of conditions. As a matter of example, we note that changing the sign of the second-harmonic dispersion term $\eta_2$ in Figs.~\ref{FigMI} and~\ref{simple_MI} leads to the disappearance of the MI bands around $d\approx 0$. This can be explained by noting that, when $\mathrm{sgn}[\eta_2]\neq\mathrm{sgn}[\Delta_2]$, the second-harmonic resonance condition can be fulfilled even for $d = 0$, leading to the disruption of sideband oscillation.  Finally, we wish to stress that, although the analyses presented in this section agree very well with results derived from the more general coupled mean-field Eqs.~\eqref{MF1} and~\eqref{MF2}, the simplified mean-field Eq.~\eqref{MF} as well as the associated MI gain Eq.~\eqref{MIgain} result from an approximation. Therefore all predictions of Eqs.~\eqref{MF} and~\eqref{MIgain} should be corroborated by carefully comparing the pertinent results against those obtained from Eqs.~\eqref{MF1} and ~\eqref{MF2}, as we have done here.

\section{Discussion and conclusion}
Let us briefly discuss the observability of frequency combs in doubly resonant intracavity SHG, considering the system in~\cite{ricciardi_frequency_2015} as an illustrative example. This configuration can be made doubly resonant by replacing all of the cavity mirrors with ones that exhibit high reflectivity both at the fundamental wavelength of 1064~nm as well as at the second-harmonic of 532~nm. Moreover, independent control of the two detunings can be implemented by managing the intracavity dispersion, e.g. through an adjustable dispersive plate. It is reasonable to assume that the cavity losses at the fundamental and the SH wavelengths are approximately equal, such that $\alpha = \alpha_2/\alpha_1 = 1$. The $L = 15~\mathrm{mm}$ length of the lithium niobate crystal, combined with the linear dispersion coefficients of ${k}''_{1} \approx 0.234~\mathrm{ps^2m^{-1}}$, ${k}''_{2} \approx 0.714~\mathrm{ps^2m^{-1}}$, and $\Delta k' \approx 792~\mathrm{ps\,m^{-1}}$, yield $\eta_1 = 1$, $\eta_2 = 3$ and $d\approx 2000$. The SHG interaction coefficient $\kappa \approx 11.4~\mathrm{W^{-1/2}m^{-1}}$ and cavity finesse $\mathcal{F}_1 =\pi/\alpha_1 = 160$, such that for critical coupling ($\alpha_1 = \theta_1$), $S = 5$ corresponds to a driving power $P_\mathrm{in} \approx 6.5~\mathrm{mW}$---much smaller than the power level used in~\cite{ricciardi_frequency_2015}. For these attainable parameters, our calculations predict MI with a frequency shift as high as 1~THz.

In conclusion, we have theoretically examined the temporal and spectral dynamics of doubly resonant intracavity SHG. We have found that the system can be accurately modelled by using two coupled mean-field equations, provided that the walk-off between the fundamental and the second-harmonic fields is appropriately included in the analysis. Our work indeed shows that, similarly to the case of singly resonant cavity SHG~\cite{leo_walkoff_2015}, the walk-off plays a key role in the dynamics. For example, in addition to MI identified previously for the case of negligible walk-off, an instability with very different characteristics arises for large walk-off values that are typical of realistic systems. Numerical simulations carried out in this regime predict that cavity MI may lead to the formation of stable (and unstable) pulse trains and corresponding optical frequency combs. These predictions give examples of the rich nonlinear behaviour that can be anticipated, but we emphasize that a considerable body of future work is needed to fully explore the solutions and their bifurcations across different parameter planes. (For example we have found that the walk-off highly impacts the Hopf bifurcation reported in~\cite{etrich_pattern_1997,etrich_solitary_1997}.) We have also derived, for the first time to our knowledge, a single reduced mean-field equation that captures the main dynamics in doubly resonant SHG. This framework has enabled us to obtain a simple expression for the system's MI gain characteristics, allowing for deep insights into the underlying physics.

We expect that our work will have significant predictive impact on the emerging field of frequency comb generation in $\chi^{(2)}$ microresonators, where several fields are likely to resonate simultaneously. Moreover, while we have here restricted our attention to the case of doubly resonant cavity-enhanced SHG, the model equations can be straightforwardly adapted to describe a variety of other dispersive, quadratically nonlinear cavity processes. These include cavities where second- and third-order nonlinearities both contribute to the dynamics~\cite{miller_onchip_2014,jung_green_2014}, sum-frequency generation in resonators~\cite{strekalov_optical_2014} as well as optical parametric oscillators~\cite{furst_low_2010, beckmann_highly_2011, werner_blue-pumped_2012}. Finally, we note that quadratically nonlinear resonators have historically attracted significant interest in the field of nonlinear dynamics and pattern formation~\cite{oppo_formation_1994, oppo_spatiotemporal_1994,santagiustina_two_1998, lodahl_pattern_1999, lodahl_pattern_2000, lodahl_modification_2000}, and we expect that the novel dynamical regimes identified in our work will also stimulate substantial further research in that context.

\section*{Funding Information}
Marsden Fund of The Royal Society of New Zealand, Rutherford Discovery Fellowships of The Royal Society of New Zealand, the Swedish Research Council (grant no. 2013-7508), the Finnish Cultural Foundation and the Italian Ministry of University and Research (grant no. 2012BFNWZ2 and Progetto Premiale QUANTOM -- Quantum Opto-Mechanics).

\appendix
\section{Derivation of mean-field equations}
\label{Apx1}
Here we derive the coupled mean-field equations~\eqref{MF1} and~\eqref{MF2}, starting from the infinite-dimensional cavity map discussed in~\cite{leo_walkoff_2015} (see also~\cite{buryak_optical_2002}). Specifically, we consider slowly varying electric field envelopes $A(z,\tau)$ and $B(z,\tau)$ with carrier frequencies $\omega_0$ and $2\omega_0$, respectively, circulating in a dispersive, quadratically nonlinear ring resonator that is driven with a cw field $A_\mathrm{in}$ at the fundamental frequency $\omega_0$. The intracavity fields $A_{m+1}(0,\tau)$ and $B_{m+1}(0,\tau)$ at the beginning of the $(m+1)^{\mathrm{th}}$ round trip can be related to the fields $A_{m}(L,\tau)$ and $B_{m}(L,\tau)$ at the end of the $m^{\mathrm{th}}$ round trip as:
\begin{align}
A_{m+1}(0,\tau) &= \sqrt{1-\theta_1}A_{m}(L,\tau)e^{-i\delta_1}+\sqrt{\theta_1}A_\mathrm{in}\label{boundary_fun}\\
B_{m+1}(0,\tau) &= \sqrt{1-\theta_2}B_{m}(L,\tau)e^{-i\delta_2}.\label{boundary_SH}
\end{align}
Here, $\theta_1$ and $\theta_2$ are the power transmission coefficients at $\omega_0$ and $2\omega_0$, respectively, of the coupler used to inject the cw field $A_\mathrm{in}$ into the resonator, $L$ is the length of the nonlinear medium and $\delta_{1}$ ($\delta_2$) represents the phase detuning of the intracavity field $A$ ($B$) from the cavity resonance closest to $\omega_0$ ($2\omega_0$).

The evolution of the fields over a single pass through the nonlinear medium obeys the coupled equations~\cite{menyuk_solitary_1994,buryak_optical_2002}
\begin{align}
&\hspace{-3pt}\frac{\partial A_m}{\partial z} =\left[-\frac{\alpha_{c1}}{2}- i\frac{{k}_1''}{2}\frac{\partial^2}{\partial \tau^2}\right]\hspace{-2pt} A_m+i\kappa B_mA_m^*e^{-i \Delta k z}, \label{fundamental_K}\\
&\hspace{-3pt}\frac{\partial B_m}{\partial z} =\left[-\frac{\alpha_{c2}}{2} - \Delta {k}'\frac{\partial }{\partial \tau}-i\frac{{k}_2''}{2}\frac{\partial^2 }{\partial \tau^2}\right]\hspace{-2pt} B_m+i\kappa A_m^2 e^{i\Delta k z},\label{SH_K}
\end{align}
where $\alpha_{c1,2}$ describe propagation losses, $\Delta k = 2k(\omega_0)-k(2\omega_0)$ is the wave-vector mismatch associated with the SHG process $\omega_0+\omega_0 = 2\omega_0$, $\Delta {k}' = \mathrm{d}k/\mathrm{d}\omega|_{2\omega_0}-\mathrm{d}k/\mathrm{d}\omega|_{\omega_0}$ describes temporal walk-off arising from the corresponding group-velocity mismatch, and ${k}''_{1,2} = \mathrm{d}^2k/\mathrm{d}\omega^2|_{\omega_0, 2\omega_0}$ are the group-velocity dispersion coefficients. (Including dispersion effects beyond the second-order is straightforward, but omitted for simplicity.) The nonlinear coupling strength is described by $\kappa\propto \chi^{(2)}$, normalised such that $|A_m|^2$, $|B_m|^2$  and $|A_\mathrm{in}|^2$ are all measured in watts (see ~\cite{leo_walkoff_2015} for details). It is noteworthy that \eqref{fundamental_K} and \eqref{SH_K} are both written in a reference frame that is moving at the group velocity of light at the fundamental frequency $\omega_0$.

In contrast to the singly resonant system examined in~\cite{leo_walkoff_2015}, here we are interested in resonators that exhibit low losses around both the fundamental and the second-harmonic frequencies. Moreover, we assume that the combined effects of nonlinearity and dispersion are ``weak'', such that the fields $A(z,\tau)$ and $B(z,\tau)$ remain approximately constant over one round trip. Under these conditions, the infinite dimensional map described by Eqs.~\eqref{boundary_fun}--\eqref{SH_K} can be reduced to two coupled mean-field equations. Following usual techniques~\cite{haelterman_dissipative_1992}, we first integrate ~\eqref{fundamental_K} and \eqref{SH_K} from $z=0$ to $z = L$, with $A(z,\tau)$ and $B(z,\tau)$ constant along $z$:
\begin{widetext}
 \begin{align}
\hspace{-3pt}A_m(L,\tau) &\approx A_m(0,\tau) + L\left[-\frac{\alpha_{c1}}{2}- i\frac{{k}_1''}{2}\frac{\partial^2}{\partial \tau^2}\right]\hspace{-2pt} A_m(0,\tau) 
+i\kappa L B_m(0,\tau)A_m^*(0,\tau)e^{-i \xi}\mathrm{sinc}(\xi), \label{fundamentalI_K}\\
\hspace{-3pt}B_m(L,\tau) &\approx B_m(0,\tau)+ L\left[-\frac{\alpha_{c2}}{2} - \Delta {k}'\frac{\partial }{\partial \tau}-i\frac{{k}_2''}{2}\frac{\partial^2 }{\partial \tau^2}\right]\hspace{-2pt} B_m(0,\tau) 
+i\kappa L A_m^2(0,\tau)  e^{i\xi}\mathrm{sinc}(\xi),\label{SHI_K}
\end{align}
\end{widetext}
where $\xi = \Delta k L /2$. We then substitute~\eqref{fundamentalI_K} into~\eqref{boundary_fun} and~\eqref{SHI_K} into~\eqref{boundary_SH}. Considering the cavity detunings $\delta_{1,2}$, out-coupling coefficients $\theta_{1,2}$, and the nonlinear and dispersive terms in~\eqref{fundamentalI_K} and~\eqref{SHI_K} as quantities of the first order and by neglecting higher-order terms, we obtain:
\begin{widetext}
 \begin{align}
A_{m+1}(0,\tau)-A_m(0,\tau) &= \left[ -\alpha_1-i\delta_1- i\frac{{k}_1''L}{2}\frac{\partial^2}{\partial \tau^2} \right]A_m(0,\tau)
+i\kappa L B_m(0,\tau)A_m^*(0,\tau)e^{-i \xi}\mathrm{sinc}(\xi) + \sqrt{\theta_1}A_\mathrm{in}, \label{MF0_F} \\
B_{m+1}(0,\tau)-B_m(0,\tau) &= \left[ -\alpha_2-i\delta_2- \Delta {k}'L\frac{\partial }{\partial \tau}- i\frac{{k}_2''L}{2}\frac{\partial^2}{\partial \tau^2} \right]B_m(0,\tau)
+i\kappa L A_m^2(0,\tau)e^{i \xi}\mathrm{sinc}(\xi) \label{MF0_S},
\end{align}
\end{widetext}
\noindent where $\alpha_{1,2} = (\theta_{1,2}+\alpha_{c1,2}L)/2$. We then introduce a continuous ``slow time'' variable $t$ that describes the evolution of the fields after each pass at $z = 0$. Defining $A(t,\tau)$ and $B(t,\tau)$ as the the cavity field envelopes at $z = 0$, we are able to replace the round-trip index $m$ with the slow time variable:
\begin{align}
A(t=mt_\mathrm{R},\tau) &= A_m(z=0,\tau), \hspace{10pt} m = 0, 1, 2, ... \label{F_cond}\\
B(t=mt_\mathrm{R},\tau) &= B_m(z=0,\tau), \hspace{10pt} m = 0, 1, 2, ... \label{SH_cond}
\end{align}
Here $t_\mathrm{R}$ corresponds to the cavity round-trip time at the reference frame where \eqref{fundamental_K} and \eqref{SH_K} have been derived, i.e., at the fundamental frequency $\omega_0$. Finally, approximating the finite-difference $\left[A_{m+1}(z=0,\tau)-A_{m}(z=0,\tau)\right]/t_\mathrm{R}$ with the partial derivative $\partial A/\partial t$ (and performing a similar approximation for the second-harmonic fields), we obtain from~\eqref{MF0_F} and~\eqref{MF0_S} the following coupled mean-field equations:
 \begin{align}
t_\mathrm{R}\frac{\partial A}{\partial t} &= \left[ -\alpha_1-i\delta_1- i\frac{{k}_1''L}{2}\frac{\partial^2}{\partial \tau^2} \right]A \nonumber \\
&+i\kappa L BA^*e^{-i \xi}\mathrm{sinc}(\xi) + \sqrt{\theta_1}A_\mathrm{in}, \label{MF_F} \\
t_\mathrm{R}\frac{\partial B}{\partial t} &= \left[ -\alpha_2-i\delta_2- \Delta {k}'L\frac{\partial }{\partial \tau}- i\frac{{k}_2''L}{2}\frac{\partial^2}{\partial \tau^2} \right]B \nonumber \\
&+i\kappa L A^2e^{i \xi}\mathrm{sinc}(\xi) \label{MF_S}.
\end{align}
With the normalization described in Section~\ref{models}, we recover Eqs.~\eqref{MF1} and~\eqref{MF2}.

For completeness, we next present the derivation of the reduced mean-field Eq.~\eqref{MF} in dimensional form. By assuming the second-harmonic field to evolve slowly, such that $\partial B/\partial t \approx 0$, we obtain from \eqref{MF_S}:
\begin{equation}
\left[ \alpha_2+i\delta_2+ \Delta {k}'L\frac{\partial }{\partial \tau}+ i\frac{{k}_2''L}{2}\frac{\partial^2}{\partial \tau^2} \right]B = i\kappa L A^2e^{i \xi}\mathrm{sinc}(\xi).
\end{equation}
We can easily find a solution for $B$ in the Fourier domain:
\begin{align}
\mathscr{F}[B] &= \frac{i\kappa L \mathscr{F}\left[A^2\right]e^{i \xi}\mathrm{sinc}(\xi)}{\alpha_2+i\delta_2 - i\Delta {k}'L\Omega - i\frac{{k}_2''L}{2}\Omega^2}\\
& = i\kappa Le^{i \xi}\mathrm{sinc}(\xi) \mathscr{F}\left[A^2\right]\hat{J}_\mathrm{d}(\Omega) \label{simpler},
\end{align}
where $\mathscr{F}\left[\cdot\right]=\int_{-\infty}^{\infty}\cdot\,e^{i\Omega\tau}\,\mathrm{d}\tau$ denotes Fourier transformation and we defined the frequency-domain nonlinear response function
\begin{equation}
\hat{J}_\mathrm{d}(\Omega) = \frac{1}{\alpha_2+i\delta_2 - i\Delta {k}'L\Omega - i\frac{{k}_2''L}{2}\Omega^2}.
\end{equation}
Substituting \eqref{simpler} into \eqref{MF_F}, we obtain
\begin{align}
t_\mathrm{R}\frac{\partial A}{\partial t} &= \left[ -\alpha_1-i\delta_1- i\frac{{k}_1''L}{2}\frac{\partial^2}{\partial \tau^2} \right]A \nonumber\\
&-(\kappa L)^2 \mathrm{sinc}^2(\xi) A^* \left[A^2 \otimes J_\mathrm{d}\right] + \sqrt{\theta_1}A_\mathrm{in},
\end{align}
where $J_\mathrm{d} = \mathscr{F}^{-1}[\hat{J}_\mathrm{d}]$. This dimensional equation is formally identical with the reduced mean-field equation presented and examined in section~\ref{Reduced}.

\section{Cavity detunings in natural phase-matching}
\label{Apx2}
The cavity phase detunings $\delta_{1,2}$ are formally defined through the equation $\phi_{1,2} =  2\pi l_{1,2} - \delta_{1,2}$, where $l_{1,2}$ is the order of the cavity resonance closest to $\omega_0$ and $2\omega_0$, respectively, and $\phi_{1,2}$ is the propagation phase accumulated by the corresponding intra-cavity fields over one round trip. Assuming, as we do in section III and IV, a monolithic (micro)ring resonator that is made entirely out of the nonlinear medium, we have $\phi_1 = k(\omega_0)L$ and $\phi_2 = k(2\omega_0)L$. For natural phase matching, $k(2\omega_0) = 2k(\omega_0)$, such that $\phi_2 = 2k(\omega_0)L = 2(2\pi l_1 - \delta_1) = 4\pi l_1-2\delta_1$. Comparison with the formal definition of $\phi_2$ makes clear that $l_2 = 2l_1$ and that $\delta_2 = 2\delta_1$. By applying the normalization $\Delta_{1,2} = \delta_{1,2}/\alpha_1$, we thus obtain the relationship $\Delta_2 = 2\Delta_1$ introduced in Section~\ref{MIsec}.

\section{Modulation instability analysis of the full cavity map}
\label{Apx3}
Here we outline the method that allows the linear stability of the infinite-dimensional map described by Eqs.~\eqref{boundary_fun}--\eqref{SH_K} to be analysed. The general principle is similar to that used in the analysis of Kerr nonlinear cavities~\cite{mclaughlin_new_1985,coen_modulational_1997, hansson_frequency_2015}.

To simplify notation, we assume phase-matching, i.e., $\Delta k = 0$. We start from the following ansatz:
\begin{eqnarray}
A(z) &=& A_0(z) + a_1(z){e}^{{i}\Omega\tau}+ b_{1}(z){e}^{-{i}\Omega\tau}\\
B(z) &=& B_0(z) + a_2(z){e}^{{i}\Omega\tau}+ b_{2}(z){e}^{-{i}\Omega\tau},
\end{eqnarray}
where $A_0(z)$ and $B_0(z)$ correspond to the mixed-mode cw ($\partial/\partial\tau = 0)$ steady-state solutions of the system described by Eqs.~\eqref{boundary_fun}--\eqref{SH_K}. In general, the steady-state fields $A_0(z)$ and $B_0(z)$ evolve as they propagate over a single round trip, and their precise evolution profiles must be obtained from numerical simulations of Eqs.~\eqref{boundary_fun}--\eqref{SH_K}. By injecting the above ansatz into Eqs.~\eqref{fundamental_K} and \eqref{SH_K}, and by linearising the resulting set of equations with respect to $a_i, b_i$, we obtain the following system of ordinary differential equations~\cite{ferro_periodical_1995}:
\begin{equation}\label{eqMI}
\frac{d\vec{a}(z)}{dz} = \boldsymbol{M}(z)\vec{a}(z),
\end{equation}
where the vector $\vec{a} = \left(a_1, b_1^*,a_2, b_2^*\right)^\mathrm{T}$ and the non-zero elements of the $4\times4$ matrix $\boldsymbol{M} := (M_{l,k}) $ are defined as
\begin{eqnarray}
\small
\begin{array}{l l} \nonumber
M_{1,1} = \displaystyle\frac{i}{2} {k}_1''\Omega^2-\displaystyle\frac{1}{2}\alpha_{c1} & M_{1,2} =i\kappa B_{0}(z) \\
M_{1,3} = i\kappa A_{0}^*(z) & M_{2,1} = -i\kappa B_{0}^*(z) \\
M_{2,2} = -\displaystyle\frac{i}{2} {{k}}_1''\Omega^2-\displaystyle\frac{1}{2}\alpha_{c1} & M_{2,4} = -i\kappa A_{0}(z)\\
M_{3,1} = 2i\kappa A_{0}(z) & M_{3,3} = -i\Delta {{k}}'\Omega + \displaystyle\frac{i}{2} {{k}}_2''\Omega^2-\displaystyle\frac{1}{2}\alpha_{c2} \\
M_{4,2} = -2i\kappa A_{0}^*(z) & M_{4,4} = -i\Delta {{k}}'\Omega - \displaystyle\frac{i}{2} {{k}}_2''\Omega^2-\displaystyle\frac{1}{2}\alpha_{c2}.
\end{array}
\end{eqnarray}

During the $m^\mathrm{th}$ round trip in the cavity, the general solution of Eq.~\eqref{eqMI} can be written as
\begin{equation}\label{general}
\vec{a}^{(m)}(z) = \boldsymbol{\boldsymbol{\Phi}}(z)\vec{a}^{(m)}(0),
\end{equation}
where $\boldsymbol{\Phi}(z)$ is the principal fundamental matrix solution of Eq.~\eqref{eqMI}, with $\boldsymbol{\Phi}(0) = \boldsymbol{I}_4$. Specifically, $\boldsymbol{\Phi}(z) = [\vec{w_1}(z), \vec{w_2}(z), \vec{w_3}(z), \vec{w_4}(z)]$, where the column vectors $\vec{w_i}(z)$ are the linearly independent solutions of Eq.~\eqref{eqMI} with initial conditions $\vec{w_1}(0) = (1, 0, 0, 0)^{\mathrm{T}}$, $\vec{w_2}(0) = (0, 1, 0, 0)^{\mathrm{T}}$, $\vec{w_3}(0) = (0, 0, 1, 0)^{\mathrm{T}}$ and $\vec{w_4}(0) = (0, 0, 0, 1)^{\mathrm{T}}$. They are obtained by numerically integrating Eq.~\eqref{eqMI}.

Equation~\eqref{general} only describes the evolution of the sideband perturbations over a single round trip. To gauge whether net growth (i.e., MI) occurs from round trip to round trip, also the cavity boundary conditions must be considered. To this end, we first write the boundary conditions given by Eq.~\eqref{boundary_fun} and \eqref{boundary_SH} for the sideband perturbations as
\begin{equation}
\vec{a}^{(m+1)}(0) = \boldsymbol{K}\vec{a}^{(m)}(L),
\end{equation}
where $\boldsymbol{K} = \mathrm{diag}(K_1,K_1^*,K_2,K_2^*)$ is a $4\times4$ diagonal matrix whose elements $K_{1,2} = \sqrt{T_{1,2}}\mathrm{exp}(-i\delta_{1,2})$ with $T_{1,2} = 1-\theta_{1,2}$.  By substituting $\vec{a}^{(m)}(L)$ from the general solution given by Eq.~\eqref{general}, we obtain the system
\begin{align}
\vec{a}^{(m+1)}(0) &= \boldsymbol{K}\boldsymbol{\Phi}(L)\vec{a}^{(m)}(0), \nonumber \\
& = \boldsymbol{Q}\vec{a}^{(m)}(0),\label{final}
\end{align}
where we introduced the matrix $\boldsymbol{Q} = \boldsymbol{K}\boldsymbol{\Phi}(L)$. By expressing $\vec{a}^{\,(m)}(0)$ and $\vec{a}^{\,(m+1)}(0)$ as a linear superposition of the eigenvectors of $\boldsymbol{Q}$, we see easily that MI is governed by the corresponding eigenvalues. Specifically, if any of the eigenvalues of $\boldsymbol{Q}$ has a modulus higher than unity, then the sideband perturbation grows from round trip to round trip, indicating MI.

To quantitatively compare the MI gain predicted by the analysis above with that obtained from the mean-field approach, we first note that, for all the cases that we considered, only one of the eigenvalues of $\boldsymbol{Q}$ has modulus larger than unity. In this case, it is easy to see that, for $m\gg 1$, one may approximate $\vec{a}^{\,(m)}(0)\approx \vec{\nu}$, where $\vec{\nu}$ is an eigenvector corresponding to the unstable eigenvalue. Substituting this approximation into Eq.~\eqref{final}, we obtain
 \begin{equation}\label{cm}
\vec{a}^{\,(m+1)}(0) = q\vec{a}^{\,(m)}(0),
\end{equation}
where $q$ is the (unstable) eigenvalue corresponding to the eigenvector $\vec{\nu}$. In the mean-field limit, the intra-cavity fields do not exhibit considerable evolution over a single round trip, implying $|q|\approx 1+\epsilon$, with $\epsilon>0$ small. Equation~\eqref{cm} can then be written as
 \begin{equation}
\vec{a}^{\,(m+1)}(0) \approx e^{|q|-1}e^{i\phi}\vec{a}^{\,(m)}(0),
\end{equation}
where $\phi = \mathrm{arg}(q)$. Referring to the dimensionless slow-time scale $t = \alpha_1t_\mathrm{d}/t_\mathrm{R}$, where $t_\mathrm{d}$ is its dimensional counterpart, we thus find that the amplitude gain $G(t) = \mathrm{exp}\left[(|q|-1)t/\alpha_1\right]$. The expression $(|q|-1)/\alpha_1$ can therefore be directly compared with the real part of the normalized eigenvalue $\lambda$ obtained from the mean-field approach in Section~\ref{MIsec} (see ansatz $v_k$). By calculating the eigenvalue $q$ for dimensional variables that correspond to the normalized quantities listed in the caption of Fig.~\ref{FigMI}, we find indeed that the expression $(|q|-1)/\alpha_1$ gives almost identical results to those shown in Fig.~\ref{FigMI}(a). This agreement is explicitly illustrated in Fig.~\ref{FigMI}(c), where we compare the spectral MI gain profiles obtained from the two approaches for two different values of the walk-off parameters ($d = 0$ and $d = 150$). For completeness, we list the dimensional variables used in the full map MI analysis: $\kappa = 2.5~\mathrm{W^{-1/2}m^{-1}}$; $L = 30~\mathrm{cm}$; $k_1'' = -20~\mathrm{ps^2km^{-1}}$; $k_2'' = -10~\mathrm{ps^2km^{-1}}$; $P_\mathrm{in} = 45~\mathrm{mW}$; $\theta_1 = 0.1$. Moreover, we assume critical coupling such that $\alpha_{c1} = \theta_1/L$. Finally, $\alpha_{c2} = \alpha_{c1}/2$ and $\theta_{2} = \theta_{1}/2$, ensuring that $\alpha = 0.5$, as in Fig.~\ref{FigMI}(a).

\end{document}